\documentclass[twocolumn,english,round,groupedaddress,superscriptaddress,reprint]{revtex4-2}
\usepackage[T1]{fontenc}
\usepackage[latin9]{inputenc}
\usepackage{fancyhdr}
\usepackage{babel}
\usepackage{varwidth}
\usepackage{amsmath}
\usepackage{amssymb}
\usepackage{graphicx}
\usepackage[bookmarks=true,bookmarksnumbered=false,bookmarksopen=false,
 breaklinks=false,pdfborder={0 0 1},backref=false,colorlinks=false]
 {hyperref}
\hypersetup{pdftitle={Can ZnO  Transform Cavity-QED?},
 pdfauthor={Wance Wang, Dhruv Fomra, Amit Agrawal, Henry J. Lezec, Joseph W. Britton},
 pdfkeywords={transparent conductive oxide (TCO), indium tin oxide (ITO), zinc oxide (ZnO), cavity quantum electrodynamics (cQED), quantum networking, quantum computing, atomic ion qubits, Rydberg atom qubits}}

\makeatletter

\newcommand*\LyXZeroWidthSpace{\hspace{0pt}}
\providecommand{\tabularnewline}{\\}
\newenvironment{cellvarwidth}[1][t]
    {\begin{varwidth}[#1]{\linewidth}}
    {\@finalstrut\@arstrutbox\end{varwidth}}

\newcommand{\bra}[1]{\left<{#1}\right|}
\newcommand{\ket}[1]{\left|{#1}\right>}
\newcommand{\braket}[3]{$\left<{#1}|{#2}|{#3}\right>$}

\newcommand{\ybp}[1]{^{#1}\mbox{Yb}^+}
\global\long\def\sn#1#2#3{^{#1}\mbox{#2}_{#3}}%

\makeatother

\begin{document}
\global\long\def\sn#1#2#3{^{#1}\mbox{#2}_{#3}}%

\global\long\def\ybp#1{^{#1}\text{Yb}^{+}}%

\global\long\def\ion#1#2{^{#1}\mbox{#2}^{+}}%

\global\long\def\nunit#1#2{#1\ #2}%

\global\long\def\bra#1{\left<#1\right|}%

\global\long\def\ket#1{\left|#1\right>}%

\global\long\def\braket#1#2#3{\left<#1|#2|#3\right>}%

\global\long\def\mmu#1{\mbox{ }\mu\mbox{#1}}%

\global\long\def\us{\text{ }\mu\text{s}}%

\global\long\def\chem#1#2{\mbox{#1}_{\mbox{#2}}}%

\author{Wance Wang}
\affiliation{Department of Physics, University of Maryland, College Park, Maryland, USA}
\affiliation{Institute for Research in Electronics and Applied Physics, University of Maryland, College Park, Maryland, USA}
\author{Dhruv Fomra}
\affiliation{National Institute of Standards and Technology, Gaithersburg, Maryland, USA}
\author{Amit Agrawal}
\affiliation{DEVCOM Army Research Laboratory, Adelphi, Maryland, USA}
\affiliation{Department of Engineering, University of Cambridge, Cambridge, UK}
\author{Henri J. Lezec}
\affiliation{National Institute of Standards and Technology, Gaithersburg, Maryland, USA}
\author{Joseph W. Britton}
\affiliation{Department of Physics, University of Maryland, College Park, Maryland, USA}
\affiliation{DEVCOM Army Research Laboratory, Adelphi, Maryland, USA}
\title{Can TCOs Transform Cavity-QED?}
\begin{abstract}
Transparent conductive oxides (TCO) enable confinement of charge-sensitive
ions and Rydberg atoms proximal to dielectric structures including
waveguides and photon detectors. However, optical loss precludes the
use of TCOs within high-finesse optical micro-resonators. Here we
characterize a ZnO-based TCO that markedly reduces optical absorption.
At $1650\text{ nm}$ we observe a $22,000$ finesse in a Fabry-Pérot
optical cavity coated with a $30\text{ nm}$ ZnO layer. This is a
5000 times reduction relative to indium tin oxide (ITO) at this wavelength.
The same ZnO film exhibits $0.01\text{ \ensuremath{\Omega}\ensuremath{\cdot}cm}$
surface resistivity at DC. We anticipate a step change in cavity-QED
systems incorporating ultra-low loss TCOs like ZnO.
\end{abstract}

\maketitle

\section{Introduction}

The coupling between atoms and electromagnetic fields underpins many
physical phenomena. Cavity quantum electrodynamics (cQED) concerns
the limit where atoms interact strongly with the modes of a resonant
structure \citep{kimble1998strong-coupling,raimond-haroche2001rmpa}.
Hallmarks of this limit include suppression \citep{gabrielse-dehmelt1985prla,heinzen-feld1987prla}
and enhancement \citep{goy-haroche1983prla,heinzen-feld1987prla}
of spontaneous emission, single atom lasing \citep{an-feld1994prla,mckeever-kimble2003na}
and generation of many-atom entangled states \citep{mcconnell-vuletic2015na}.
Today, cQED systems appear in variety of contexts including quantum
networking \citep{Krutyanskiy23prl,teller-northup2023aqsa,Reiserer22rmp},
quantum simulation \citep{dordevic-lukin2021s,Sauerwein23np,Clark20n,Mivehvar21ap},
precision measurement and sensing \citep{Greve22n,Li23s,Robinson24np}.
Control of atomic systems using microfabricated photonic structures
permits on-chip optical addressing \citep{Niffenegger20n,mehta2020nature-integrated-optical-qip,hsu-regal2022pqa}
and tailored optical potentials \citep{Keil16jmo,Zhou24prx}, but
these techniques have had limited impact on cQED with charge-sensitive
atoms and ions. If surface-charging could be mitigated there are big
wins attainable from miniaturization since atom-cavity coupling $g_{0}$
scales with mode volume $V$ as $g_{0}\sim V^{-1/2}$.

Photo-electron accumulation on dielectric surfaces proximal to ions
confined in RF Paul traps complicates trapping and laser cooling \citep{berkeland-wineland1998micromotion,harlander-blatt2010charging,wang-chuang2011joapa,brown-haffner2021nrm}.
And it's a leading technical challenge when integrating optical cavities
with ions \citep{brandstatter-northup2013rsia,bruzewicz-sage2019apr,takahashi2020ion-strong-coupling,Ong20njp,teller-northup2023aqsa}
and Rydberg atoms \citep{saffman-saffman2016jpbamop,ocola-lukin2024prl}.
Surface charging has consequences in other domains as well. For example,
photon sources based on quantum dots (QD) in optical cavities suffer
from spectral wandering due to surface charges \citep{somaschi-senellart2016np,senellart-white2017nna,liu-fox2018nn}
and mirror charging is problematic for LIGO \citep{pollack-gundlach2010prda,martynov-zweizig2016prd}.
Of course, a thin metal layer can block stray electric fields but
this spoils the high reflectivity of super-cavity mirrors \citep{takahashi2020ion-strong-coupling,teller-northup2023aqsa}.

Transparent conductive oxides (TCOs) strike a compromise between the
optical transparency of a good insulator and the electrical conductivity
of a metal. Indium tin oxide (ITO) is a widely used TCO when high
single-pass optical loss is acceptable. It has been applied to shield
trapped ions from integrated photonics and detectors in surface-electrode
traps \citep{Niffenegger20n,eltony-chuang2013apla} and from free-space
large NA lenses \citep{Carter24rosi}. However, ITO exhibits high
optical loss due to free-carrier absorption and scattering in the
near-infrared \citep{Wang05apl,Wang18ci,Wang19pssa,Maniyara21am}.

Optical properties of a lossy dielectric can described by the complex
refractive index $\tilde{n}=n+i\kappa$. The term $4\pi\kappa/\lambda$
is the exponential decay constant for light with wavelength $\lambda$.
To get a feel for the problem note that a single $10\text{ nm}$-thick
intracavity film with $\kappa=0.04$ at $1550\text{ nm}$ limits a
Fabry--Pérot cavity finesse to $\sim1000$. At $1550\text{ nm}$,
the typical $\kappa$ for ITO is greater than $0.5$ \citep{Koida12em,Xian19oe}.
At this wavelength $\kappa\sim0.2$ for other TCOs like niobium- and
fluorine-doped tin oxide \citep{Su13cap,Biedrzycki14jpcc,Ching-Prado18jmsme}.

In this paper we discuss the ZnO material system, detail how we optimize
a growth recipe and then characterize its optical and electrical properties.
Our recipe yields a ZnO thin film with $\kappa\sim10^{-4}$ at $1650$
nm ($\sim10\text{ ppm}$) while simultaneously exhibiting a surface
resistivity of $\sim0.01\text{ \ensuremath{\Omega}\ensuremath{\cdot}cm}$.
In Sec.$\:$\ref{sec:stray-charge} we show this resistivity is sufficient
to mitigate photoelectric charging issues for both ions and Rydberg
atoms in compact optical cavities. This development is immediately
impactful to ongoing work exploiting telecom-wavelength transitions
in ion \citep{Wang20oq2c2pq,Luo09fp} and neutral atom \citep{Huie21prr,Craddock24pra,thompson2024rydberg-interconnect}
systems. We expect similar low loss below $1000\text{ nm}$ which
would have high impact \citep{saffman-saffman2016jpbamop,Krutyanskiy23prl,Evered23n}
including mid-circuit readout in Rydberg-based quantum computers \citep{Deist22prl,ocola-lukin2024prl}.

\section{zno material system }

\label{sec:material}

ZnO, a II--IV semiconductor, is notable for its intrinsic n\nobreakdash-type
behavior, high carrier concentration, low electron effective mass,
high electron mobility, low electron affinity and exceptional radiation
hardness \citep{borysiewicz2019mdpi-review,ozgur2005ajp-review}.
Its direct bandgap of $\sim3.3\text{ eV}$ suggests negligible absorption
in the visible and near\nobreakdash-infrared spectrum \citep{hala2015tco-nir,Wang18ci,Wang19pssa,Spencer22apr}.
Applications of ZnO thin films include suppression of field emission
\citep{lv-zhai2018ci}, use as a transparent thin-film conductor in
solar cells \citep{hala2015tco-nir,Alonso-Alvarez17se}, use as a
thin-film transistor in high-power electronics \citep{yang-ye2023aaem}
and more recently as an epsilon-near-zero NIR nonlinear optical material
\citep{kinsey-boltasseva2015o,secondo-kinsey2022apl,ball-khurgin2023ainp,Fomra24apr}.

At room temperature the thermodynamically stable crystalline structure
is wurtzite B4. For a wide range of growth conditions wurtzite ZnO
behaves as an intrinsic n-type semiconductor with very high electron
densities \citep{ozgur2005ajp-review}. Leading explanations for
the high electron density in the absence of explicit doping include
intrinsic point defects (self-interstitial Zn type $\mathrm{Zn_{i}}$
and $\mathrm{O_{2}}$ vacancy type $\mathrm{O_{v}}$) and contamination
by hydrogen during deposition \citep{mccluskey-jokela2009joap,vidya-nieminen2011prb,kohan-vandewalle2000prb}.
In an Argon-rich Oxygen-poor sputtering environment with a ZnO target,
$\mathrm{O_{v}}$ are prevalent due to the higher yield of Zn relative
to $\mathrm{O_{2}}$ \footnote{There is a notable disparity in the energy transfer factor between
$\text{Ar-Zn}=0.94$ and $\text{Ar-O}=0.81$ for RF sputter deposition.}\citep{lin-lii2004sact}. However, recent studies suggest $\mathrm{O_{v}}$
are deep-donor defects calling into question their contribution to
n-type conductivity \citep{ozgur2005ajp-review,walle2000H2inZnO}.
Hydrogen interstitials on the other hand are unquestioned as shallow,
positive charge state donors. The hydrogen forms a strong, stable
bond with oxygen contributing to the wide range of conditions under
which n-type conductivity is observed \citep{ozgur2005ajp-review}.
Hydrogen contamination is ubiquitous in stainless steel deposition
chambers and is likely incorporated during film growth \citep{walle2000H2inZnO}.

Across the visible spectrum optical absorption in many reported ZnO
thin films is low, but documented performance in the near-IR is surprisingly
poor given its small band gap \citep{hala2015tco-nir}. Contributing
factors include non-stoichiometric composition that could cause free
carrier absorption, sub-band gap absorption-like defects and large
Urbach tailing \citep{rai-rai2013joap,fomra-kinsey2020ome,wang-xu2020oe}.
This suggests several paths to reduce near-IR absorption: improved
control over film stoichometry \citep{Singh15jap}, attention to contamination
in the growth chamber and annealing to reduce defects.

Al-doped ZnO (AZO) and Ga-doped ZnO are promising TCOs for applications
like in solar cells \citep{hala2015tco-nir,Alonso-Alvarez17se} and
epsilon-near-zero nonlinear optics \citep{Wang19pssa,Fomra24apr}.
However, their documented optical transmission of AZO in the NIR is
poor \citep{Wang18ci} -- we did not explore approaches to optimize
it for ultra-low loss.

\section{process development}

The NIST microfabrication facility has an established process for
sputter deposition of ZnO thin films that served as a baseline for
process development. We studied two growth parameters as a function
of near-IR absorption and DC electrical conductivity: stoichiometry
and annealing.

Thin films of ZnO were deposited on two substrate types: fused silica
and thermally oxidized silicon. The substrates were cleaned with acetone
and IPA. The deposition was carried out using RF magnetron sputtering
(Denton Vacuum, Discovery 550\citep{no-endorsement}), employing
a 3-inch diameter high-purity ($99.999\%$) ZnO ceramic target. Prior
to deposition, the sputtering system was evacuated to a high vacuum
of approximately $5\times10^{-7}$ Torr using a turbo molecular pump.
The sputter target-to-substrate separation was $10\text{ cm}$. Prior
to deposition the target was cleaned by pre-sputtering for $120\text{ s}$
with the shutter closed. During each fixed-duration $300\text{ s}$
deposition cycle, the system configuration was as follows. Process
gas flowed into the chamber at $50\text{ cm}^{3}/\text{min}$ and
the pressure was maintained at $5\times10^{-3}\text{ Torr}$. The
gas composition was a variable process parameter; it included Argon,
Oxygen and Nitrogen. The RF power was held constant at $170\text{ W}$
at $13.56\text{ MHz}$ and was coupled to the plasma by an automatic
LC matching network. Neither the substrate nor the sputter target
temperature was actively controlled. The deposition rate was not actively
controlled.

\begin{figure*}
\includegraphics[width=0.7\paperwidth]{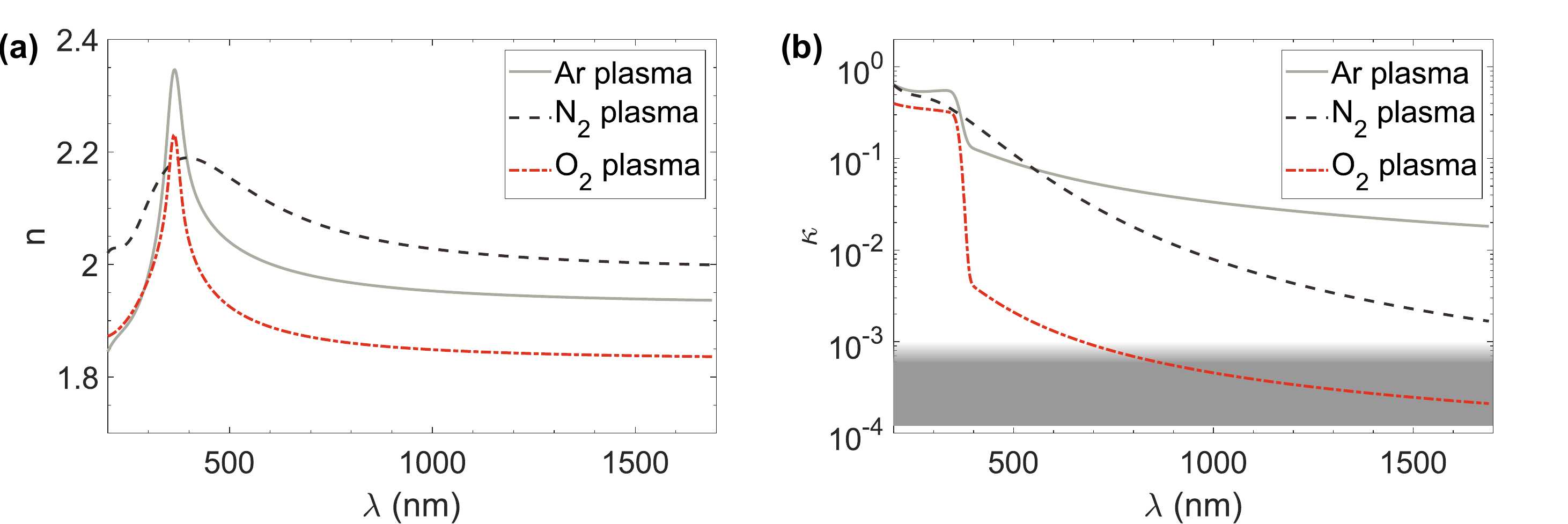}\caption{\protect\label{fig:index-wavelength}Three thin film samples were
sputtered with $100\text{\%}$ Ar, $\mathrm{N_{2}}$ and $\mathrm{O_{2}}$.
Ellipsometer measurements indicate (a) index of refraction and (b)
optical absorption. The accuracy of the ellipsometer is poor for $\kappa<10^{-3}$
; this region is shaded grey.}
\end{figure*}

Post-deposition, the thin film broadband optical properties were evaluated
using an ellipsometer (M2000, JA Woollam \citep{no-endorsement} (see
Fig.$\:$\ref{fig:index-wavelength} and Appendix$\:$\ref{subsec:ellipsometer-model})
and the resistivity was measured using the Hall Effect and a 4-point
probe (see Appendix \ref{subsec:electrical-properties}).

\begin{figure*}
\includegraphics[width=0.7\paperwidth]{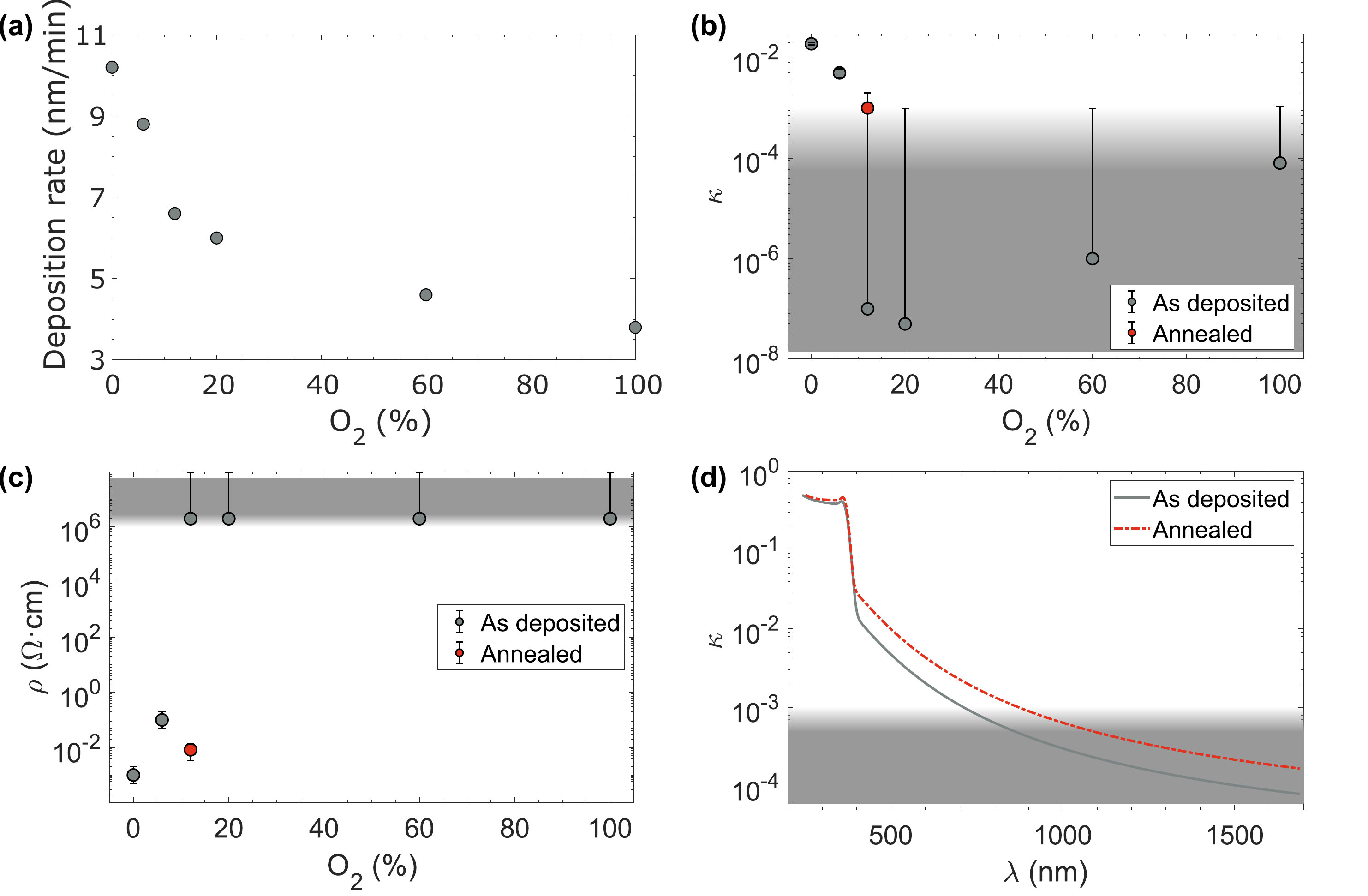}\caption{\protect\label{fig:loss-resitivity}Thin films were sputtered with
Argon-Oxygen with a variable Oxygen fraction. The plots show the (a)
deposition rate, (b) optical loss at $1550\text{ nm}$ and (c) resistivity.
One sample was annealed; its post-annealing measurements are marked
in red in (b) and (c). Plot (d) compares the optical loss of an annealed
and an unannealed thin film. Some thin film samples exhibited properties
outside the reliable measurement range of the ellipsometer and 4-point
probe. The unreliable ranges are delineated by shaded grey.}
\end{figure*}

We anticipate that an excess of free electrons dominates optical loss
in the NIR due to Drude scattering. This suggests biasing our search
toward higher resistance films. Empirically it is observed that depositing
ZnO in $\text{O}_{2}$ rich conditions reduces the free carrier density
\citep{damiani-mansano2012jpcs} and leads to a more stochiometric
film \citep{damiani-mansano2012jpcs}. To explore this route, several
ZnO thin films were sputtered with different process gas mixtures.
We used a gas mixture composed of Ar and $\mathrm{O_{2}}$ with varying
oxygen fraction. Increasing $\mathrm{O_{2}}$ fraction leads to a
reduction of deposition rate, a reduction in optical absorption and
an increase in resistivity (see Fig.$\:$\ref{fig:loss-resitivity}a-c).
Therefore, the first step in our process development was growing thin
films with three process gas configurations: 1) 100\% $\mathrm{O_{2}}$,
2) 100\% $\mathrm{N_{2}}$ and 3) 100\% Ar. The films sputtered with
$100\text{\%}$ oxygen exhibited the lowest optical losses in the
near-IR ($1300\text{ nm}$ to $1650\text{ nm}$) but exhibited very
high resistivity. The films sputtered with argon and nitrogen exhibited
relatively larger optical loss and significantly lower resistivities.
We use these process extremes to bracket our subsequent process optimization.
Fig.$\:$\ref{fig:index-wavelength} shows the measured index of refraction
for the several process gas configurations.

Engineered approaches to hydrogen doping include H-ion implantation
\citep{kennedy-reeves2006cap}, H-plasma exposure \citep{dhakal-westgate2012jovs&ta},
introduction of hydrogen during growth \citep{zhu-xie2013jovs&ta}
and post-deposition annealing in forming gas or hydrogen gas \citep{jiang-zhao2013ml}.
Motivated by this we annealed several ZnO samples in a forming gas
consisting of $97\text{\%}\text{ }\mathrm{N_{2}}$ and $3\text{\% }\mathrm{H_{2}}$
for $300\text{ s}$ ($10\mathrm{^{\circ}C/min}$ ramp rate). The temperature
was limited to $400\mathrm{^{\circ}C}$ to avoid etching the ZnO,
known to occur at $\sim500\mathrm{^{\circ}C}$ \citep{yin-wang2016aa}.
We annealed samples and found that the resistivity dropped from $>10^{6}\text{ \ensuremath{\Omega}\ensuremath{\cdot}cm}$
to $\sim10\text{ m\ensuremath{\Omega}\ensuremath{\cdot}cm}$. In
all cases the resistivity decreased but the change in optical loss
was hard to track as it was outside the range of the ellipsometer
(Fig. \ref{fig:loss-resitivity}). As a control we annealed sample
films in a $\mathrm{N_{2}}$ environment and saw no change in resistivity.

Based on these observations, we settled on a recipe as detailed above:
a gas mixture with $12\text{\% }\mathrm{O_{2}}$ is used during sputtering,
and a process time is selected to achieve a ZnO thickness $h=30\text{ nm}\pm2\text{ nm}$--the
error is due to the ellipsometer-based thickness measurement. We then
prepared ZnO films on several high-finesse mirrors and proceeded with
further characterization. 

\section{Optical characterization\protect\label{subsec:Optical-Properties}}

Measuring small optical losses is challenging. The accuracy of the
ellipsometry tool is poor for $\kappa<10^{-3}$. A system exquisitely
sensitive to optical loss is the Fabry-Perot optical cavity (FPI).
For a cavity with finesse $\mathcal{F}=10^{4}$ the circulating light
bounces $\sim10,000$ times before exiting the cavity. A small change
in cavity mirror reflectivity can be measured by noting the change
in cavity ring-down time. Salient FPI cavity properties are its transmission
frequency $\nu_{c}$, the resonance full width at half max linewidth
$\delta\nu_{c}$ and free spectral range (FSR) $\nu_{\mathrm{FSR}}$.
The finesse is $\mathcal{F}=\nu_{\mathrm{FSR}}/\delta\nu_{c}$. The
approach we use to measure $\kappa$ is to build a high-finesse FPI
test cavity and measure the ring-down time, then coat one mirror with
ZnO and repeat the ring-down measurement.

Our test cavity consists of a pair of mirrors; one is planar and the
other has radius of curvature $R=-50\text{ mm}$. The mirror substrate
is super-polished fused silica with surface roughness $<0.84\:\mathring{\mathrm{A}}\,\mathrm{rms}$
and a highly reflective optical coating deposited by ion-beam sputtering
by FiveNine Inc \citep{no-endorsement}. The coating consists of alternating
layers of $\mathrm{Ta_{2}O_{5}}$ and $\mathrm{SiO_{2}}$ with $\mathrm{Ta_{2}O_{5}}$
as the final layer. The coating transmission measured by the manufacturer
is reported to be $\sim0.012\%$ at $1650\text{ nm}$. The substrates
and coatings are tolerant to annealing at up to $480\mathrm{^{\circ}C}$.

\begin{figure*}
\includegraphics[width=0.6\paperwidth]{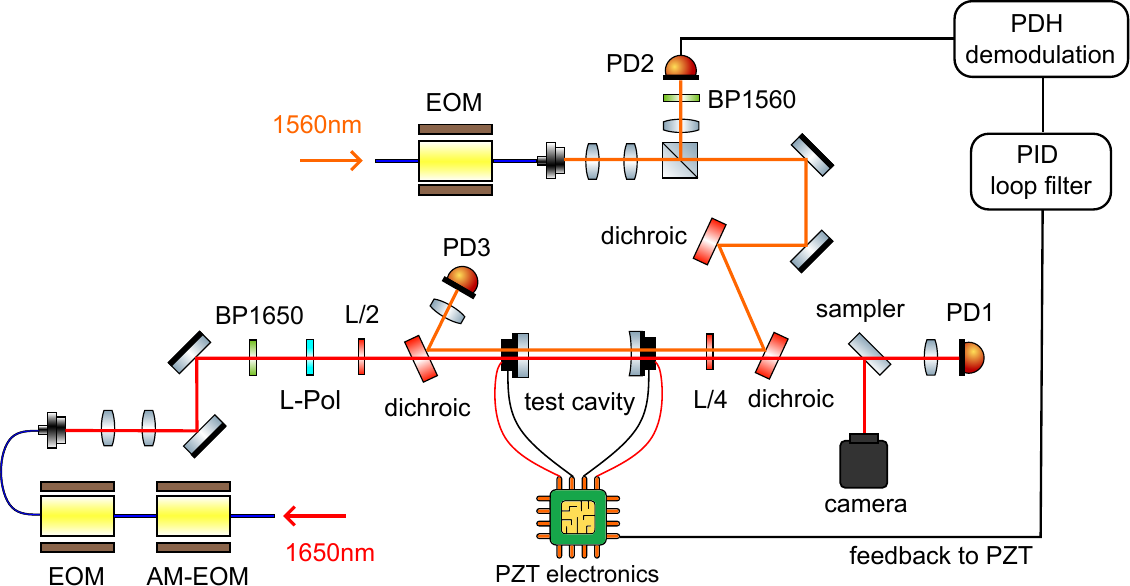}\caption{\protect\label{fig:fpi-setup}The optical and feedback setup of the
ring-down experiment on a length-stabilized test cavity.}
\end{figure*}

Figure \ref{fig:fpi-setup} shows the ring-down setup. The test cavity
mirror spacing $d=20.2\text{ mm}$ is controlled by PZT to keep the
cavity on resonance with a laser at $1560\text{ nm}$. The Pound-Drever
Hall (PDH) technique is used to actively servo $d$ to enforce this
condition. A second laser at $1650\text{ nm}$ is tuned to be simultaneously
resonant with the cavity. Both lasers are coupled to the fundamental
transverse cavity mode $\text{TEM}_{00}$ with a contrast exceeding
$15:1$. And, both lasers are frequency stabilized to a common ultra-stable
reference cavity (Stable Laser System, Inc \citep{no-endorsement})
using the PDH offset-locking technique with a frequency uncertainty
$\ll\delta\nu_{c}$ \citep{Wang24arX}.

We measured linewidth $\delta\nu_{c}$ and FSR $\nu_{\mathrm{FSR}}$
to calculate the finesse $\mathcal{F}=\nu_{\mathrm{FSR}}/\delta\nu_{c}$
at $1650\text{ nm}$. The linewidth was measured by ring-down measurements
as follows. A photodetector (PD) measured the power transmitted by
the cavity. A wideband EOM amplitude modulator (AM-EOM, iXblue MXER-LN-10
\citep{no-endorsement}) was used to rapidly attenuate the laser light.
A measurement cycle consisted of the steps: lock the cavity length
to the $1560\text{ nm}$ laser, then rapidly extinguish the laser
light with the AM-EOM and record the decaying PD voltage signal $V_{PD}$.
An exponential fit $V_{PD}=V_{0}e^{-2\pi\delta\nu_{c}t}$ to a multiple
successive ring-down traces yielded fitting parameters $\delta\nu_{c}$
and $V_{0}$. The cavity FSR was determined by measuring the average
frequency splitting between adjacent cavity transmission features
using a near-IR wave meter (Bristol 228A \citep{no-endorsement}).
Since modifying the cavity mirror coating requires temporary mirror
removal (which could change $d$), we repeat the $\nu_{\mathrm{FSR}}$
measurement after each reassembly.

The finesse of a cavity with two mirrors with reflectivity $r_{i}$
and $r_{j}$ is $\mathcal{F}_{ij}=\pi\sqrt{r_{i}r_{j}}/(1-r_{i}r_{j})$,
where $r_{i}$ is the change in electric field amplitude $E_{0}r_{i}$
after a single reflection. Let $\mathcal{F}_{00}$ be the cavity finesse
with identical mirrors $r_{0}$ and $r_{0}$; let $\mathcal{F}_{01}$
be the cavity finesse where one mirror is $r_{0}$ and the other,
coated by the thin film or changed by annealing, is $r_{1}$.
\begin{align}
r_{0} & =\frac{\sqrt{4\mathcal{F}_{00}^{2}+\pi^{2}}-\pi}{2\mathcal{F}_{00}}\label{eq:refl_1}\\
r_{1} & =\frac{1}{r_{0}}\left(\frac{\sqrt{4\mathcal{F}_{01}^{2}+\pi^{2}}-\pi}{2\mathcal{F}_{01}}\right)^{2}\label{eq:refl_2}
\end{align}
The attenuation of light in a lossy medium decays as $E_{0}e^{-\frac{2\pi}{\lambda}\kappa z}$
and the corresponding complex index of refraction is $\tilde{n}=n+i\kappa$.
Now we can see how an FPI can measure the extinction coefficient of
a thin film applied to a cavity mirror. The transmittance plus fractional
power loss for the mirror with $r_{i}$ is $1-r_{i}^{2}$. The excess
loss is $r_{0}^{2}-r_{1}^{2}$ when the reflectivity changes from
$r_{0}$ to $r_{1}$, assuming the transmittance is unchanged by the
thin film. We ascribe any excess loss to absorption in the thin film,
$r_{0}^{2}-r_{1}^{2}=1-e^{-\frac{4\pi}{\lambda}\kappa2h}$, where
$h$ is the film thickness. Thus we get
\begin{equation}
\kappa=-\frac{\lambda}{8\pi h}\ln\left(1-r_{0}^{2}+r_{1}^{2}\right)\label{eq:extinction-coeff}
\end{equation}

Table$\:$\ref{tab:loss-date} summarizes the three FPI cavity configurations
for which we measured the finesse and extinction coefficients. After
annealing and exposure to air for 4 months we observed a $15\%$ increase
in finesse. As ZnO is strongly polar and chemically active \citep{borysiewicz2019mdpi-review},
we speculate that the improvement may be due to adsorbed atmospheric
gases and is compatible with reports in the literature \citep{hala2015tco-nir,Radjehi23mme}
 Our results show a ultra low loss of ZnO layer in a cavity with
finesse $>10^{4}$. Since in our calculation all excess loss was considered
absorption loss, the actual absorption loss could be lower.

The uncertainty in $\mathcal{F}$ is dominated by statistical uncertainty
in $\delta\nu_{c}$. As an example, the uncertainties underlying the
finesse observation made after $27$ days are $\delta\nu_{c}=(523\pm9)\text{ kHz}$
and $\nu_{\mathrm{FSR}}=(7.410\pm0.013)\text{ GHz}$. We have good
confidence that neither noise or systematic effects complicate interpretation
of the trends in Table$\:$\ref{tab:loss-date}.

\begin{table*}
(a) %
\begin{tabular}{|c|c|c|c|c|}
\hline 
finesse & before deposition & 27 days & 69 days & 128 days\tabularnewline
\hline 
\hline 
$M_{0}:M_{0}$ & $23,340\pm60$ & / & / & /\tabularnewline
\hline 
$M_{0}:M_{\mathrm{ZnO}}$ & / & $14,160\pm250$ & $18,400\pm700$ & $19,800\pm180$\tabularnewline
\hline 
$M_{0}:M_{A}$ & / & / & $20,900\pm300$ & $22,120\pm130$\tabularnewline
\hline 
\end{tabular}

(b) %
\begin{tabular}{|c|c|c|c|}
\hline 
$\kappa/10^{-5}$ & 27 days & 69 days & 128 days\tabularnewline
\hline 
\hline 
$M_{\mathrm{ZnO}}$ & $38\pm3$ & $16\pm3$ & $8.0\pm0.7$\tabularnewline
\hline 
$M_{A}$ & / & $6.7\pm1.1$ & $3.2\pm0.4$\tabularnewline
\hline 
\end{tabular}

\caption{\protect\label{tab:loss-date}The cavity finesse was measured for
three mirror configurations: two symmetric bare mirrors (denoted $M_{0}:M_{0}$),
one bare mirror and one ZnO-coated mirror (denoted $M_{0}:M_{\mathrm{ZnO}}$),
and one bare mirror and one bare but annealed mirror (denoted $M_{0}:M_{A}$).
Measurements were performed at different intervals following the initial
deposition. Table (a) is the observed finesse. The uncertainty is
one standard deviation. Table (b) lists the extinction coefficient
$\kappa$ calculated from Eq.$\:$\ref{eq:extinction-coeff} where
$r_{0},r_{1}$ are solved using Eq.$\:$\ref{eq:refl_1} and \ref{eq:refl_2}.
The thickness $h=(30\pm2)\text{ nm}$ is determined by ellipsometry
and the test wavelength is $\lambda=1650\text{ nm}$. The uncertainty
includes contributions from $\mathcal{F}$ and $h$.}
\end{table*}

In addition to ringdown measurements, we verified that depositing
ZnO film on the existing high-reflector stack does not significantly
alter the mirror's power transmission $T$. The mirror's power reflection
$R$ drops from $r_{0}^{2}$ to $r_{1}^{2}$, which we attribute to
film absorption $A$, while assuming $T$ remains unchanged. Although
in principle $R+T+A=1$, our measurements of $T$ via the reflection
dip on cavity resonance \citep{Rempe92ol} showed no substantial variation.
At 61 and 128 days after deposition, we measured $T=(1.21\pm0.05)\times10^{-4}$
and $(1.37\pm0.02)\times10^{-4}$, respectively. Compared to the vendor-specified
$T=1.18\times10^{-4}$, the variation is at most $1.9\times10^{-5}$.
In contrast, the observed reflection variations $r_{0}^{2}-r_{1}^{2}$
are $7.2\times10^{-5}$ and $4.8\times10^{-5}$ at 69 and 128 days.
Thus, we conclude that the change in transmission is not dominant.

The refractive index of ZnO at $1650\text{ nm}$ is approximately
1.9 (see Fig.$\,$\ref{fig:index-wavelength}(a)). High-finesse optical
coatings typically use alternating layers of $\mathrm{Ta_{2}O_{5}}$
and $\mathrm{SiO_{2}}$, with indices of about $2.1$ and $1.5$ at
$1650\text{ nm}$. Incorporating ZnO layers into the multi\nobreakdash-layer
stack design can help minimize any changes in transmission.

\section{Electrical characterization}

\label{subsec:electrical-properties}

Electrical properties were measured using a Hall Effect system and
a 4-point probe. For these measurements thin films were sputtered
on glass and thermal $\text{SiO}_{2}$ squares using the recipe described
above. The film thickness was determined by ellipsometry; typical
thickness was $h\sim10\text{ nm}$. The results are in Table$\:$\ref{tab:resistivity}.

For the Hall Effect measurement, the samples were squares ($L_{H}=25\text{ mm}$)
cut from the sputtered wafers. The Hall probe contact diameter was
$D_{H}\approx10\text{ \ensuremath{\mu}m}$. In the limit $L_{H}\gg h,D_{H}$,
the Hall measurement is reliable \citep{nist-fab-hall-measurement}.
For the 4-point probe measurements edge effects were negligible. 

\begin{table*}
\begin{tabular}{|c|c|c|c|c|c|}
\hline 
Sample & Substrate & \begin{cellvarwidth}[t]
\centering
4-point probe

($\mathrm{m\ensuremath{\Omega}\ensuremath{\cdot}cm}$)
\end{cellvarwidth} & \begin{cellvarwidth}[t]
\centering
Hall effect

($\mathrm{m\ensuremath{\Omega}\ensuremath{\cdot}cm}$)
\end{cellvarwidth} & \begin{cellvarwidth}[t]
\centering
Carrier concentration

($\mathrm{cm^{-3}}$)
\end{cellvarwidth} & \begin{cellvarwidth}[t]
\centering
Mobility

($\mathrm{cm^{2}/(V\cdot s)}$)
\end{cellvarwidth}\tabularnewline
\hline 
\hline 
ZnO-1 & glass & 8.34 & 8.6 & $2\times10^{19}$ & 37\tabularnewline
\hline 
ZnO-2 & glass & 10.1 & 13.2 & $1.5\times10^{19}$ & 28\tabularnewline
\hline 
ZnO-3 & $\mathrm{SiO_{2}}$ on Si & 8.25 & / & / & /\tabularnewline
\hline 
\end{tabular}

\caption{\protect\label{tab:resistivity}The resistivity was measured using
two techniques: 4-point probe and Hall effect. The carrier concentration
and mobility were calculated based on the resistivity.}
\end{table*}

\begin{figure}
\includegraphics[width=0.95\columnwidth]{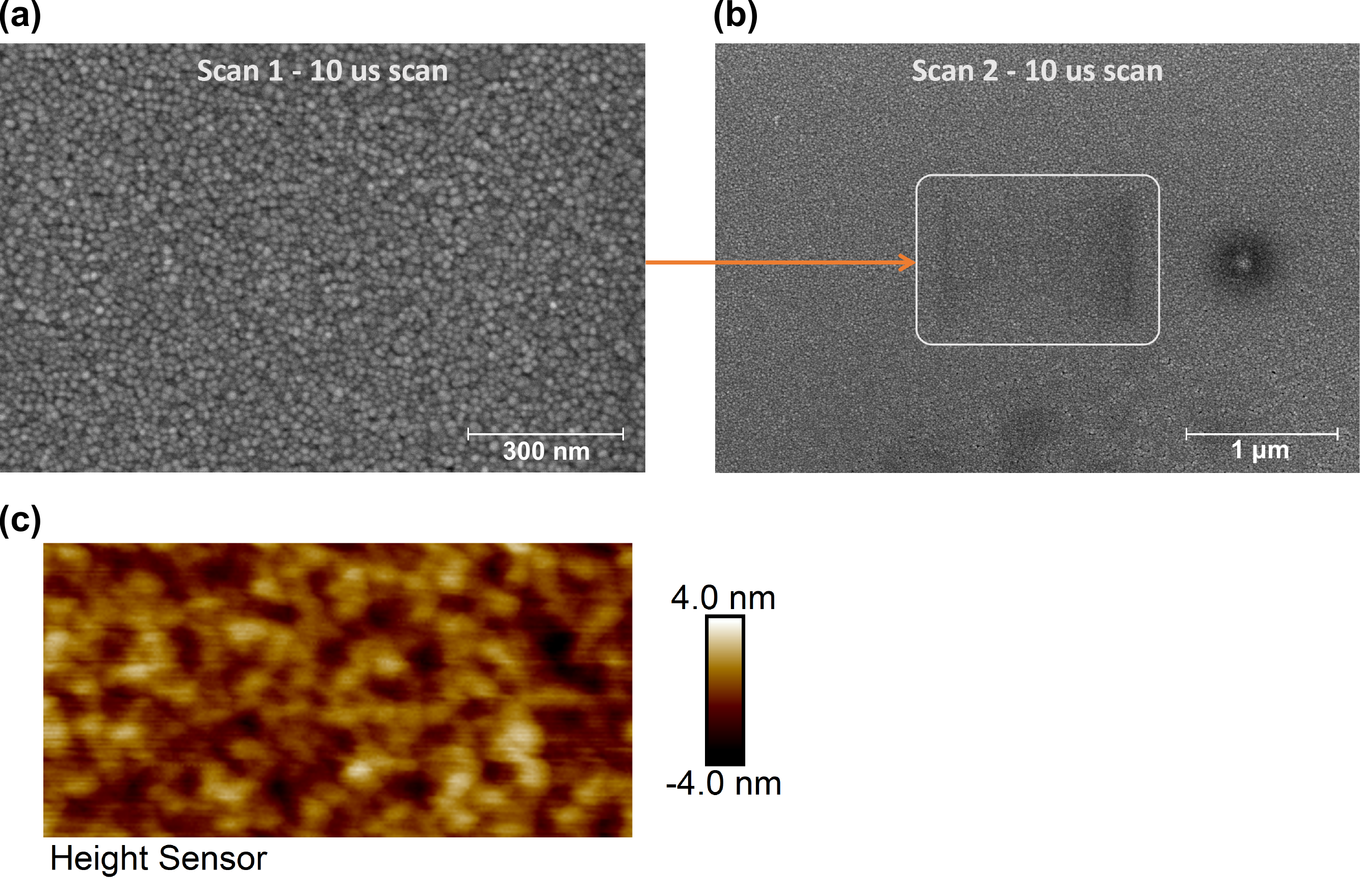}

\label{fig:surface-morphology}\caption{(a,b) Scanning electron microscope images show a uniform distribution
of spherical grains $10\text{ nm}$ to $30\text{ nm}$ in diameter.
Neither has bright spots indicative of insulating patches. The e-beam
was $5\text{ kV}$, $50\text{ pA}$ and focused to $\sim30\text{ nC/cm}^{2}$.
(c) An atomic force microscope scan showing surface roughness $<1\text{ nm}$
RMS and $3.5\text{ nm}$ peak-to-peak. }
\end{figure}

Under some deposition conditions, ZnO deposition results in complex
micro structure including nanosheets and porous structures \citep{borysiewicz2019mdpi-review}.
Further, mismatch between the thermal expansion coefficient of the
substrate and the thin film could cause surface deformation. Because
the present application calls for wide-area ($\gg\lambda$) homogeneity
in both optical phase retardation and electrical resistance, we measured
the spatial variation of surface properties using both a scanning
electron microscope (SEM) and an atomic force microscope (AFM) (see
Figure$\:$\ref{fig:surface-morphology}).

Low-resistance Ohmic contact to the ZnO film is required for most
applications. This is reportedly routine in $\text{ZnO}$ from room
temperature to $300^{\circ}\text{C}$ \citep{ozgur2005ajp-review}.

\section{stray charge and atomic systems}

\label{sec:stray-charge}

Following is a simple model for thinking about the consequences of
stray charge on optical cavity mirrors \footnote{Our model is intentionally crude. It covers key physics but omits
details needed to be prescriptive. A full analysis would consider
in detail the trap and mirror geometries, inhomogeneous charge distributions
and currents, and modification of the trapping potential due to dielectrics
(e.g., the mirrors).}. We then apply the model for trapped ions and Rydberg atoms in an
optical cavity. Finally, we determine the surface resistivity required
of a thin film to mitigate the effects of charging.

To start, let's work in one dimension. Consider a positive test charge
$q$ at position $x$ near the origin. Now add a pair of stationary
point charges $Q_{1}$ and $Q_{2}$ at locations $\pm x_{Q}$. The
interaction between the test charge and the stray charges has potential
energy
\[
\frac{q}{4\pi\epsilon_{0}}\left(\frac{Q_{1}}{\left|x+x_{Q}\right|}+\frac{Q_{2}}{\left|x-x_{Q}\right|}\right)
\]
Simplify by limiting the range of $x$ to $-x_{Q}<x<x_{Q}$ and expanding
to second order in $x$. 
\begin{equation}
U_{Q}=s_{q}\left(Ax+Bx^{2}+C\right)+\mathcal{O}(x^{3})\label{eq:U2Qo2-1}
\end{equation}
where $s_{q}=\frac{q}{4\pi\epsilon_{0}}$, $A=\frac{Q_{2}-Q_{1}}{x_{Q}^{2}}$,
$B=\frac{Q_{1}+Q_{2}}{x_{Q}^{3}}$ and $C=\frac{Q_{1}+Q_{2}}{x_{Q}}$.
The electric field is $E_{Q}(x)=-\frac{1}{q}\frac{d}{dx}U_{Q}(x)=-\frac{A+2Bx}{4\pi\epsilon_{0}}$.

Contrast this simple model with another. Consider a dielectric disc
with a surface charge density $\sigma$. When the disc diameter is
large relative to $x_{Q}$, we can model it as a sheet of infinite
extent. The potential energy of a test charge located a distance $x$
away from such a surface is $\frac{q\sigma}{2\epsilon_{0}}x$. For
a pair of surfaces at distance $\pm x_{Q}$ interacting with a test
charge $q$ near the origin, the potential energy is $q\frac{\sigma_{2}-\sigma_{1}}{2\epsilon_{0}}(x-x_{Q})$.
This corresponds to a constant electric field at the test charge.

Finally, consider a finite-diameter mirror with radius $r$ and a
uniform charge density $\sigma=Q/\pi r^{2}$. The potential energy
for interaction with a test charge $q$ at distance $x$ above the
surface is $U_{m}(x)=2s_{q}Q\frac{2\sqrt{r^{2}+x^{2}}-x}{r^{2}}$.
For comparison, the potential energy of a test charge interacting
with a point charge $Q$ is $U_{p}(x)=s_{q}Q/x$. We can now calculate
a calibration factor for a typical fiber cavity setup with $r=125\text{ \ensuremath{\mu}m}$
and $x_{Q}=200\text{ \ensuremath{\mu}m}$. We find $U_{m}/U_{p}=0.92$
and $E_{m}/E_{p}=0.78$. These values are close enough to unity to
motivate our use of this toy cavity geometry throughout the remainder
of this section.

\subsection{trapped ion sensitivity}

Let's model the implications of these stray charges when the test
charge $q$ is harmonically bound. Assume that the potential energy
is $U_{t}=k_{t}x^{2}$ where $k_{t}=\frac{1}{2}m\omega_{x}^{2}$ and
$\omega_{x}$ is the secular frequency. The total potential energy
is then $U=U_{t}+U_{Q}$. The ion equilibrium position $\tilde{x}$
is found by solving $dU/dx=0$,
\[
\tilde{x}=-\frac{1}{2}\frac{s_{q}A}{k_{t}+s_{q}B}
\]
Its eigenfrequency at position $\tilde{x}$ is
\[
\tilde{\omega}_{x}=\sqrt{\frac{1}{m}\frac{d^{2}U}{dx^{2}}|_{x=\tilde{x}}}=\sqrt{\frac{2(k_{t}+s_{q}B)}{m}}
\]
The second order effect is a secular frequency shift.

Several studies have considered the impact of stray charges on ion
trapping \citep{harlander-blatt2010charging,podoliak-horak2016praa}.
Even in the absence of charging a dielectric perturbs the trap potential
\citep{podoliak-horak2016praa} and contributes to motional heating
\citep{teller-northup2021prla,kumph-blatt2016heating-dielectric}.

\subsubsection{impact on laser cooling}

An ion displaced from the harmonic potential minimum by $\tilde{x}$
suffers excess micromotion of amplitude $x_{\mathrm{\mu m}}=\sqrt{2}\frac{\tilde{\omega}_{x}}{\Omega_{\mathrm{RF}}}\tilde{x}$
\citep{amini2011fab-traps}, where $\Omega_{\mathrm{RF}}$ is Paul
trap RF frequency. One consequence of micromotion is that in the atomic
reference frame laser light at $\lambda_{D}$ appears modulated at
the micromotion frequency $\Omega_{\mathrm{RF}}$ and the optical
carrier intensity at the ion is reduced by $J_{0}^{2}(\beta)$, where
$\beta=2\pi\frac{x_{\mathrm{\mu m}}}{\lambda_{D}}$. The scattering
rate for a transition driven below saturation scales linearly with
laser intensity. So in the case when the laser is relied upon for
Doppler cooling, the cooling power also scales as $J_{0}^{2}(\beta)$.

To gauge the scale of the problem, let the test charge be $\text{Yb}^{+}$
where $q=+e$, $m=171\text{ amu}$ and $\lambda_{D}=369\text{ nm}$,
corresponding to the cooling transition. Assume an ion trap where
$x_{Q}=200\text{ \ensuremath{\mu}m}$, $\Omega_{\mathrm{RF}}/2\pi=30\text{ MHz}$
and $\omega_{x}/2\pi=500\text{ kHz}$. To minimize the impact on laser
cooling, we require $J_{0}^{2}(\beta)<0.5$ which implies $\tilde{x}<2.8\text{ \ensuremath{\mu}m}$.
This constrains the stray charges to $Q_{1}<1400\text{ e}$ when $Q_{2}=0$.
The corresponding electric field is $E_{Q}(\tilde{x})<49\text{ V/m}$.

\subsubsection{impact on ion-cavity coupling}

Next we identify much tighter constraints on ion position in the context
of ion-cavity interactions. Suppose that $x=0$ is the location of
the cavity waist and a standing-wave antinode. For a cavity wavelength
$\lambda_{c}=1650\text{ nm}$ we calculate what charge imbalance displaces
the ion by $\lambda_{c}/8$. If $\tilde{x}=\lambda_{c}/8$ the ion-cavity
coupling drops by half. Continuing with our example, if $Q_{1}=100\text{ e}$
and $Q_{2}=0$, the resulting electric field is $E_{Q}(\tilde{x})=3.6\text{ V/m}$
and $\tilde{x}=\lambda_{c}/8$ is reached.

\subsubsection{impact on laser-mediated gates}

We consider two approaches to gauging the impact of stray charge on
the interaction of a trapped ion with an optical field of wavelength
$\lambda_{g}$.

The ion-light interaction becomes decoupled from ion motion in the
Lamb-Dicke limit: $k\left<x\right>\ll1$, where $\left<x\right>$
is the amplitude of ion motion and $k=2\pi/\lambda_{g}$. Suppose
$\lambda_{g}=355\text{ nm}$ \citep{Wang20prl} and we require $kx_{\mathrm{\mu m}}<0.2$.
This requirement is met in our example system when $\tilde{x}<0.47\text{ \ensuremath{\mu}m}$
and $x_{\mathrm{\mu m}}<11\text{ nm}$. Under these conditions and
assuming $Q_{2}=0$, the stray charge is limited to $Q_{1}<230\text{ e}$,
which corresponds to $E_{Q}(\tilde{x})<8.2\text{ V/m}$. Compare $x_{\mathrm{\mu m}}$
with the zero-point spread of $\text{Yb}^{+}$ motional wavefunction
$\sqrt{\hbar/(2m\omega_{x})}=8\text{ nm}$.

Laser-mediated 2-qubit gates are impaired by secular frequency shifts
\citep{Hayes12prl,Sutherland22pra}. Below, we rely on the fidelity
reported in Figure 2 and Equation (38) of \citep{Sutherland22pra}.
Suppose the initial spin state of two ions is $|\downarrow\downarrow\rangle$,
the gate is performed on the radial mode $\omega_{x}$, the motion
is initialized to occupation number $n_{x}$ and the two-qubit Rabi
rate is $\Omega_{2g}$. Let $\delta_{x}=|\omega_{x}-\tilde{\omega}_{x}|$
denote the motional frequency error. If $\Omega_{2g}/2\pi=10\text{ kHz}$
and $n_{x}=50$, the infidelity is smaller than $0.01$ when $\delta_{x}/\Omega_{2g}<0.013$.
In our toy $\text{Yb}^{+}$ model, if $Q_{1}=Q_{2}<630\text{ e}$,
then $\delta_{x}/\Omega_{2g}<0.013$.

\subsection{Rydberg atom sensitivity}

\subsubsection{impact on Rydberg state coherence}

Consider the $^{87}\text{Rb}$ atom in the 70S Rydberg state. Let
$E_{0}$ be the energy splitting between the ground $\left|g\right>=\left|5\text{S}_{1/2}\right>$
and $\text{\ensuremath{\left|r\right>}}=\left|70\mathrm{S_{1/2}}\right>$.
At low field the quadratic Stark effect causes a frequency shift $\delta_{R}/2\pi=\frac{1}{2}\alpha E_{Q}^{2}$,
where $\alpha=53.4\text{ kHz\ensuremath{\cdot}(V/m)}^{-2}$ is the
polarizability \citep{ocola-lukin2024prl}. This is the dominant
mechanism by which Rydberg atoms are decohered by the local charge
environment. An unknown shift $\delta_{R}$ gives rise to dephasing
$\Delta\phi$ 
\[
\Delta\phi=\delta_{R}\tau=\pi\alpha E_{Q}^{2}\tau
\]
In the context of a Ramsey experiment, when $\Delta\phi=\pi$ the
system is fully decohered which occurs after duration $\tau_{\pi}=(\alpha E_{Q}^{2})^{-1}$.
This contributes to the decoherence of entangled states \citep{Ebert15prl,ocola-lukin2024prl}
and limits cavity-based photon storage and retrieval \citep{Baur14prl,Distante17nc,Tiarks19np,Stiesdal21nc}.

A coherence times of $T_{2}^{*}\sim5\text{ \ensuremath{\mu}s}$ is
common in Rydberg atom systems \citep{levine-lukin2018prla,Saffman11jpcs,ocola-lukin2024prl}.
If $x_{Q}=200\text{ \ensuremath{\mu}m}$, $Q_{1}=54\text{ e}$ and
$Q_{2}=0$ then $E_{Q}(0)=1.9\text{ V/m}$, $\delta_{R}/2\pi=100\text{ kHz}$
and $\tau_{\pi}=5\text{ \ensuremath{\mu}s}$. 

\subsubsection{impact on gate fidelity via Rydberg blockade}

Fluctuating surface charges also reduce the fidelity of Rydberg-blockade
gates \citep{Zhang12pra,saffman-saffman2016jpbamop,DeLeseleuc18pra,levine-lukin2018prla}.
An unknown frequency shift $\delta_{R}/2\pi=\frac{1}{2}\alpha E_{Q}^{2}$
results in an incomplete excitation onto the Rydberg state with a
population $1-\delta_{R}^{2}/\Omega_{R}^{2}$, where $\Omega_{R}$
is the two-photon Rabi frequency. The resulting gate infidelity is
$1-\mathcal{F}=\frac{1}{2}\delta_{R}^{2}/\Omega_{R}^{2}$. If $\Omega_{R}/2\pi=5\text{ MHz}$
and $Q_{2}=0$, we have $1-\mathcal{F}=0.01$ when $Q_{1}=140\text{ e}$
($E_{Q}(0)=5.1\text{ V/m}$).

\section{laser-induced charging}

\begin{figure}
\includegraphics[width=0.95\columnwidth]{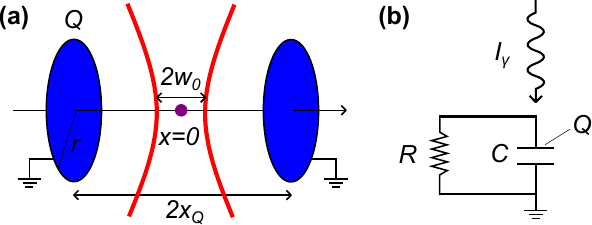}\caption{We use a simple model to estimate charging. (a) A laser beam with
a waist $w_{0}$ addresses an atom inside a Fabry--Pérot cavity formed
by TCO-coated mirrors. The wings of the Gaussian clip the mirrors
generating photocurrent $I_{\gamma}$. (b) A lump circuit model of
the TCO.}
\label{fig:charging}
\end{figure}

Physical phenomena that contribute to charge accumulation on dielectric
mirror surfaces include the photoelectric effect, cosmic rays and
radioactive decay. Here we focus on the photoelectric effect \citep{harlander-blatt2010charging,ivory-parazzoli2021yb-wg}.
In this section we estimate the sheet resistance $R_{s}$ (units $\Omega/\square$)
required to mitigate charging due to stray laser light generating
photocurrent $I_{\gamma}$ on a TCO-coated mirror (see Fig. \ref{fig:charging}).

The front face of a mirror of radius $r$ is coated by a TCO of thickness
$h$ and resistivity $\rho=R_{s}h$. The TCO is grounded at its perimeter.
We approximate that the round film's resistance as $R\approx\frac{2r}{2r}R_{s}$.
In parallel with $R$ is capacitance $C$ between the TCO and ground.
At equilibrium the voltage drop is $V=I_{\gamma}R$ and the charge
on the film is $Q=CV=RCI_{\gamma}$.

For simplicity, we make some intentionally pessimistic assumptions.
Assume $\lambda$ is below the band gap and perfect quantum efficiency
(every photon generates a photo-electron). If the laser beam power
directly incident on the mirror\footnote{Note that the tolerable laser intensity is much higher if focused
at the center of the cavity. Suppose a laser beam with waist $w_{0}=100\text{ \ensuremath{\mu}m}$
is focused on an ion at $x=0$. The light intensity at the mirrors
($x=\pm200\text{ \ensuremath{\mu}m}$) drops by $\exp\left(-x_{Q}^{2}/w_{0}^{2}\right)^{2}\approx3\times10^{-4}$.} is $0.2\text{ mW}$ and $\lambda=369\text{ nm}$ then $I_{\gamma}/\text{e}\approx4\times10^{11}\text{ s}^{-1}$.
Suppose $r=125\text{ \ensuremath{\mu}\text{m}}$, $h=30\text{ nm}$
and the TCO resistivity matches that of our ZnO recipe ($\rho=10^{-4}\text{ \ensuremath{\Omega}\ensuremath{\cdot}m}$)
so that $R=3.3\text{ k\ensuremath{\Omega}}$. Further, assume $C=0.1\text{ pF}$
\footnote{The capacitance has three contributions: a) the self capacitance of
a round conductive plate (relative to a ground at infinity) $8\varepsilon_{0}r\approx8\times10^{-3}\text{ pF}$;
b) the capacitance between the films on the two Fabry--Pérot cavity
mirrors $\epsilon_{0}\frac{\pi r^{2}}{2x_{Q}}\approx1\times10^{-3}\text{ pF}$
and c) the capacitance between the film and nearby conductors like
ion trap electrodes ( $\sim0.1\text{ pF}$). Note that a typical RC
time-constant is less than $1\text{ ns}$.}. In this scenario the equilibrium charge on the film is $Q\approx120\text{ e}$
which if unknown (or uncompensated) is nearly tolerable by the metrics
in Sec.$\,$\ref{sec:stray-charge}. Also note that the discharge
$RC$ time constant is less than $1\text{ ns}$. Since this scenario
is intentionally pessimistic we surmise that the demonstrated ZnO
resistivity suffices in many real-world scenarios of contemporary
relevance.

\section{Discussion}

Our optical characterization was performed at $1650\text{ nm}$ but
many AMO applications rely on wavelengths near $800\text{ nm}$. We
expect low material loss down to $\sim800\text{ nm}$ since the dominant
loss mechanisms are not operative. Urbach tailing and deep-level defects
produce exponential absorption near the bandgap ($\sim375\text{ nm}$)
but has little contribution at $800\text{ nm}$. And, for stoichiometric
films with low carrier density the Drude model predicts intraband
absorption scaling as $\kappa\propto\lambda^{3}$ (far above $1000\text{ nm}$\LyXZeroWidthSpace{}
\citep{Wang05apl,Singh15jap}). Neither mechanism suggests increased
loss at $800\text{ nm}$. The ellipsometer measurement (Fig \ref{fig:index-wavelength})
is suggestive of this interpretation but can't be relied upon in the
ultra-low loss regime. 

Some applications require cavity operation at cryogenic temperatures.
However, predicting the temperature dependence of TCOs is challenging
\citep{Guo11apl,Gao13ssc}. This is especially so for ZnO where there
is uncertainty about the factors contributing to its high free electron
density (see Sec.$\:$\ref{sec:material}).

This paper is an initial step toward exploring the applications of
TCOs like ZnO in ultra-low loss optics. By mitigating charging issues,
we open up new possibilities for integrating dielectric optics close
to charge-sensitive atomic systems. Future directions for ZnO include
measuring its optical properties at other wavelengths and at cryogenic
temperatures. And evaluating practical issues like its tolerance to
cleaning polymers and propensity to surface contamination (ZnO is
a strongly polar and chemically active \citep{borysiewicz2019mdpi-review}).
In the context of ion trapping, it remains to measure its resistivity
at RF frequencies which could reduce trap potential deformation due
to the mirror dielectric \citep{podoliak-horak2016praa} and could
mitigate dielectric heating \citep{kumph-blatt2016heating-dielectric,teller-northup2021prla}.

\section{Appendix}

\subsection{Ellipsometer Model}

\label{subsec:ellipsometer-model}

Variable angle spectroscopic ellipsometry (VASE) \citep{khoshman2014zno-ellip}
determines optical properties of a film under test by measuring the
wavelength-dependent change in polarization of reflected light. The
observables are the standard ellipsometric parameters: relative amplitude
change $\Psi(\lambda)$ and phase change $\Delta(\lambda)$. Models
used to interpret the data include PSemi-M0 (PSM0) oscillator, a Gaussian
oscillator and a Drude function to account for the band edge absorptions
and free electron response.

Our ZnO films were modeled using 3 oscillators: PSemi-M0, Gaussian
and Drude. The Psemi-M0 is a Kramers-Kronig consistent oscillator,
generally used to model the band edge of direct band gap semiconductors.
The Gaussian oscillator was used to account for absorption near the
band edge not captured by PSemi-M0. Lastly, for the annealed films,
a Drude term was added to model the free electron behavior. However,
due to the low free carrier density, these films do illicit a strong
Drude response in the measured NIR spectrum, thus leading to a highly
correlated resistivity and scattering term. To break this correlation,
resistivity in the optical model was fixed and set to be equal to
the electrical resistivity, even though these have been shown to deviate
due to different scattering mechanisms at play.

\begin{figure}
\includegraphics[width=0.9\columnwidth]{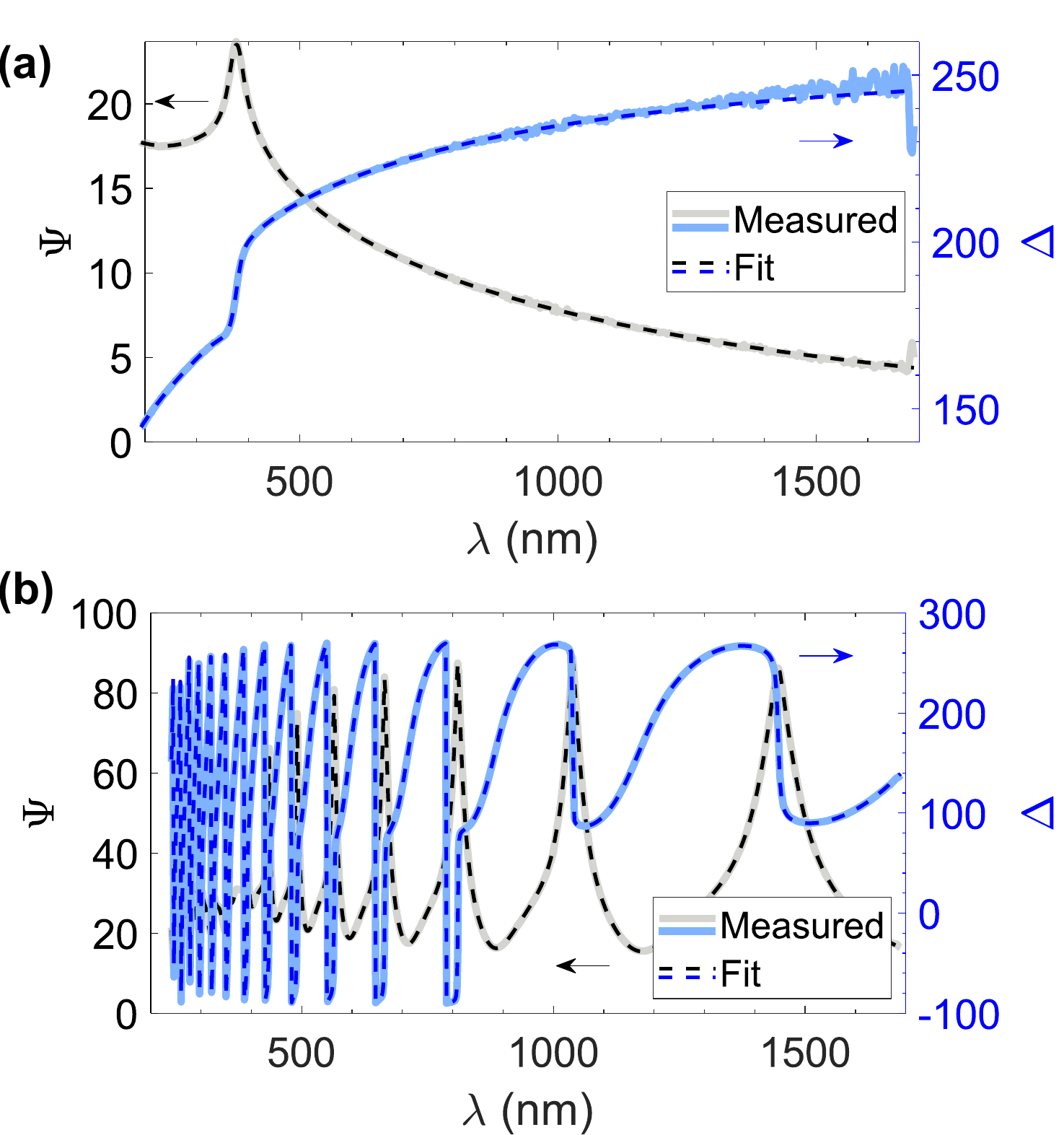}

\caption{\protect\label{fig:Ellipsometer}Plots of the measured and the modeled
$\Psi$ and $\Delta$ at a $65^{\circ}$ angle of incidence for a
$30\text{ nm}$ ZnO film deposited on (a) borosilicate glass and (b)
$1.5\text{ \ensuremath{\mu\text{m}}}$ thermal oxide on Si. The residual
is less than $5^{\circ}\text{ RMS}$.}
\end{figure}

\begin{figure}
\includegraphics[width=0.9\columnwidth]{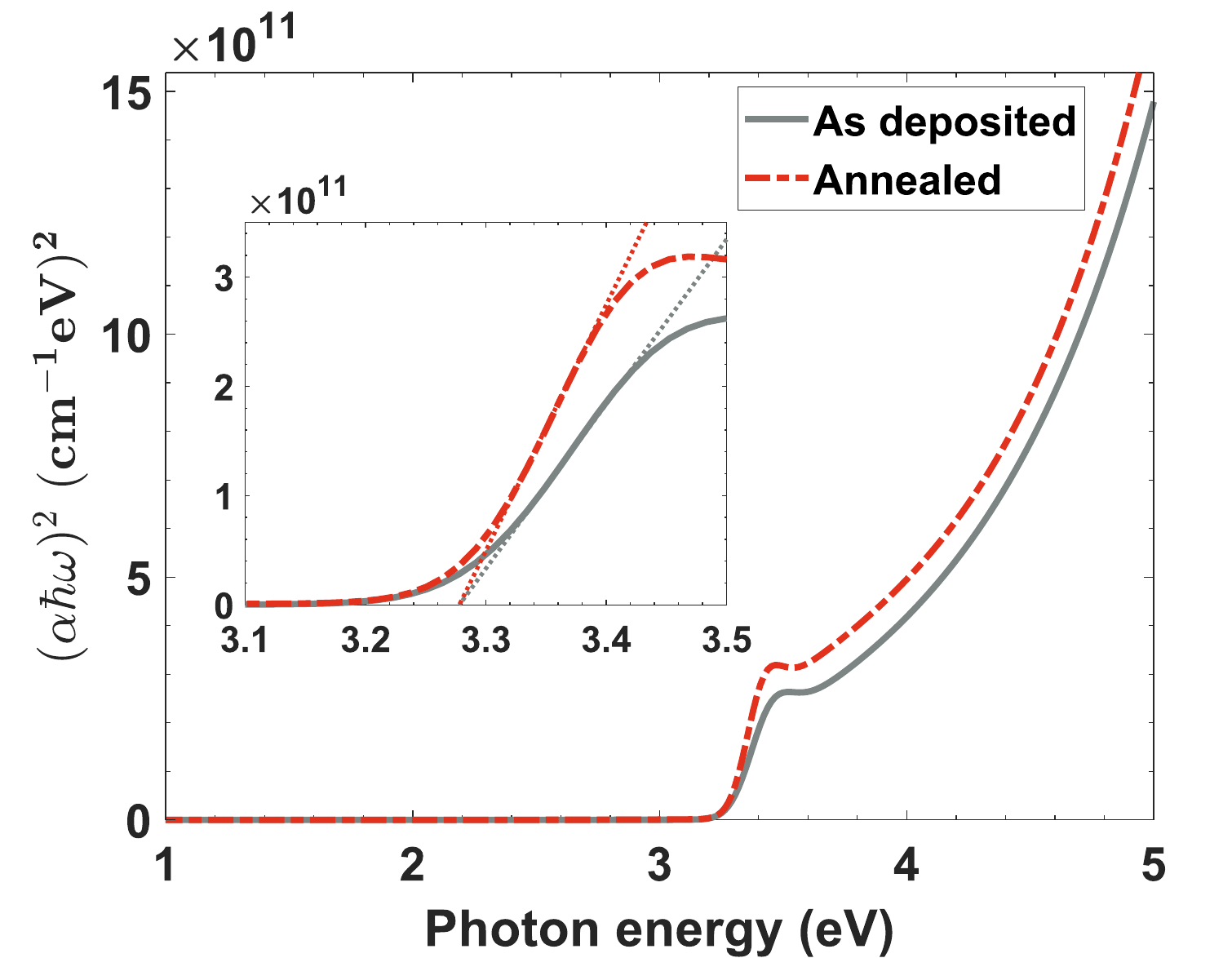}\label{fig:band-edge-estimate}

\caption{A Tauc plot of $(\alpha\hbar\omega)^{2}$ versus photon energy $\hbar\omega$
for the film deposited under 12\% $\mathrm{O_{2}}$ partial pressure,
both before and after annealing. Extrapolating the linear edge of
each curve to $(\alpha\hbar\omega)^{2}=0$ (inset) yields a band-edge
energy of $3.3\text{ eV}$ in both cases.}
\end{figure}

The fits to the $\Psi$ and $\Delta$ measurements, a ratio of s-
and p-polarized reflectivity, is shown in Figure \ref{fig:Ellipsometer}.
These measurements were made for ZnO deposited on silicon wafers.Figure
\ref{fig:band-edge-estimate} is a Tauc plot generated from the ellipsometer-measured
absorption. The extracted bandgaps before and after annealing are
virtually unchanged, indicating essentially no Moss-Burstein shift.

\section*{Bibliography}

\bibliographystyle{apsrev4-2}
\bibliography{2024tco}

\begin{thebibliography}{111}
\bibitem [{no-()}]{no-endorsement}%
  \BibitemOpen
  \href@noop {} {\bibinfo {title} {Products or companies named here are
  included in the interest of completeness and does not imply endorsement by
  the authors.}}\BibitemShut {Stop}%
\end{thebibliography}

\end{document}